\documentclass[a4paper,11pt]{article}
\pdfoutput=1 

\usepackage{jcappub} 
\usepackage{xcolor} 
\usepackage[T1]{fontenc} 
\usepackage{subcaption}

\newcommand{\DN}{\Delta N_{\textrm{eff}}}

\definecolor{darkgreen}{cmyk}{0.85,0.2,1.00,0.2}

\usepackage{ mathrsfs }

\def\nn{\nonumber}
\def\({\left(}
\def\){\right)}
\def\[{\left[}
\def\]{\right]}

\title{\boldmath BBN constraints on dark radiation isocurvature}

\author{Peter Adshead,}
\author{Gilbert Holder,}
\author{and Pranjal Ralegankar}
\affiliation{Department of Physics, University of Illinois at Urbana-Champaign, Urbana, IL 61801, USA }

\emailAdd{adshead@illinois.edu}
\emailAdd{gholder@illinois.edu}
\emailAdd{pranjal6@illinois.edu}

\abstract{The existence of  dark radiation that is completely decoupled from the standard model in the early Universe leaves open the possibility of an associated dark radiation isocurvature mode. We show that the presence of dark radiation isocurvature leads to spatial variation in the primordial abundances of helium and deuterium  due to spatial variation in $N_{\rm eff}$ during Big Bang nucleosynthesis. We use the result to constrain the existence of such an isocurvature mode on scales down to $\sim 1$ Mpc scales. By measuring the excess variance in the primordial helium to hydrogen and deuterium to hydrogen ratio in different galaxies, we constrain the variance in average isocurvature in a galaxy to be less than $0.13/\Delta \bar{N}_{\rm eff}$ at 95\% confidence. Here $\Delta \bar{N}_{\rm eff}$ is the spatially averaged increase in $N_{\rm eff}$ due to the additional dark radiation component.}

\begin{document}
\maketitle
\flushbottom
\section{Introduction}
\label{sec:intro}

Upcoming stage 4  cosmic microwave background (CMB) experiments will make exquisite measurements of the energy content of the Universe \cite{Abazajian:2016yjj}. These measurements will improve the constraint on the contribution of free-streaming radiation (through the effective number of relativistic species $N_{\rm eff}$) by an order of magnitude over current constraints.
A measurement of $N_{\rm eff}$ consistent with the standard model (SM) prediction of $N_{\rm eff} = 3.044$ \cite{Mangano_2005, Grohs_2016, deSalas:2016ztq, Akita:2020szl, Abenza_2020} will place extremely strong constraints on particle content beyond the SM \cite{Errard:2015cxa}.

Alternatively, these measurement could reveal the existence of additional light-particles (dark radiation) beyond the SM by measuring $N_{\rm eff} \neq 3.044$ at high significance. If these additional particles were ever in thermal equilibrium with the standard model, they will exhibit the usual adiabatic fluctuations in their density (see, e.g.\ \cite{Weinberg:2004kf}), and their effects on cosmology would be indistinguishable from additional neutrinos. However, this dark radiation may be completely decoupled from the standard model sector. In this decoupled scenario, fluctuations in the dark radiation density may be independent of the density fluctuations in the visible sector---there may be a dark-radiation isocurvature mode. Isocurvature modes are generically predicted by cosmological theories that have a second clock beyond the SM temperature field \cite{Lyth:2002my}. In this work we remain agnostic to the exact origin of such an initial condition and leave detailed model-building to future work.

The presence of dark radiation isocurvature affects the cosmic microwave background (CMB). Earlier work \cite{Kawasaki:2011rc} constrained neutrino + dark radiation isocurvature to be less than $\sim 10^{-5}$ at scales around 500 Mpc using data from WMAP and ACT (see also Ref.~\cite{Kawakami:2012ke} for non-Gaussian dark radiation isocurvature constraints). Planck is sensitive to dark-radiation isocurvature for scales $\geq 10$ Mpc \cite{Akrami:2018odb}.  However, the inability to observe CMB fluctuations on angular sizes smaller than $\sim$ 5 arcmins prohibits the estimation of isocurvature constraints on smaller scales \cite{Akrami:2018vks}. In this work we probe dark radiation isocurvature down to $\sim 1$~Mpc scales through its impact on Big Bang nucleosynthesis (BBN).

BBN is a period in the early universe when the SM plasma became cold enough for the free protons and neutrons to combine and form the first nuclei. This process primarily produces Helium and Deuterium (along with trace amounts of tritium and Lithium). Adiabatic fluctuations during BBN do not lead to spatial variations in the outcome of BBN. This result follows directly from  the separate universe picture---different patches of the Universe with differing density fluctuations simply appear to be a little older or younger as viewed by local observers. Since the local physics is identical, the outcome is identical. The presence of isocurvature during BBN can change the story by changing the physical conditions locally in a way that is distinguishable from a local shift in the clock. In this way, isocurvature leads to spatial variation in  primordial elemental yields. Spatial variations in the yields of ${}^4$He/H and D/H  during BBN would then lead to corresponding differences in abundances in widely separated locations. To date, baryon isocurvature modes as a source of inhomogeneous BBN have been extensively studied in the literature \cite{2010ApJ...716..907H,Inomata:2018htm,Copi:1996td,1973ApJ...179..343W,Barrow_2018}. In this work we consider the effect of dark radiation isocurvature on the primordial elemental abundances from BBN.

Our results can be summarized as follows.  We demonstrate that the presence of dark-radiation isocurvature leads to spatially varying elemental abundances. As a result, galactic ${}^4$He/H and D/H ratios are sensitive to dark-radiation isocurvature on galactic scales, $\sim 1$ Mpc. We use data on Helium abundances in nearby galaxies \cite{Cooke:2017cwo} and Deuterium abundances in high-redshift Lyman-$\alpha$ absorption systems \cite{Aver:2015iza} to place constraints on the existence of dark radiation isocurvature. We constrain the variance of  average isocurvature fluctuations in galaxies, to be less than $0.13/\Delta \bar{N}_{\rm eff}$ at $2\sigma$ confidence for scales around $\sim 1$ Mpc. Here $\Delta \bar{N}_{\rm eff}$ is the spatially averaged increase in $N_{\rm eff}$ due to the additional dark radiation component. In the absence of any dark-radiation, i.e. $\Delta\bar{N}_{\rm eff}=0$, our constraints are relaxed as expected.

This paper is organized as follows. In section \ref{sec:varying_BBN}, we show how dark-radiation isocurvature leads to spatially varying BBN yields and demonstrate that this leads to differences in the primordial abundances of light elements in different galaxies. In section \ref{sec:constraints}, we use excess variance in existing ${}^4$He/H and D/H data to place constraints on dark-radiation isocurvature. We conclude in section \ref{sec:conclusion}. Finally, in appendix~\ref{appendix:island_universe}  we use  the separate universe approach to demonstrate how dark-radiation isocurvature leads to spatially varying $\Delta N_{\rm eff}$.

\section{Inhomogenous Big Bang Nucleosynthesis through dark radiation isocurvature}\label{sec:varying_BBN}

In this section we demonstrate the impact of dark-radiation isocurvature on BBN. We first show how the effect of dark-radiation isocurvature on BBN is distinct from the more studied baryon-isocurvature case \cite{2010ApJ...716..907H,Inomata:2018htm,Copi:1996td,1973ApJ...179..343W,Barrow_2018}. We then demonstrate that dark-radiation isocurvature leads to spatially varying $N_{\rm eff}$ that in turn causes spatial variation in primordial abundances of hydrogen and helium.

The elemental abundances produced by BBN are primarily determined by two processes: the weak processes which convert neutrons to protons, and by the deuterium formation process that forms deuterium from all the remaining neutrons. The temperature at which deuterium formation begins, $T_{\rm nuc}$, is insensitive to the Hubble rate and is primarily determined by the baryon-to-photon ratio. Baryon-isocurvature modes cause the baryon-to-photon ratio to vary spatially. This in turn leads to a spatially varying $T_{\rm nuc}$, and to spatial variations of the resulting elemental abundances. In contrast, a dark-radiation isocurvature mode leads to a spatially varying $N_{\rm eff}$, as we show below. The abundance of neutrons at $T_{\rm nuc}$ is sensitive to the Hubble rate at $T_{\rm nuc}$, and since $N_{\rm eff}$ affects the Hubble rate through the Friedmann equation, an inhomogeneous $N_{\rm eff}$ leads to an inhomogeneous abundance of neutrons at $T_{\rm nuc}$.

To show how dark-radiation isocurvature leads to spatial variation in $N_{\rm eff}$, we first highlight the relation between dark-radiation energy density and $N_{\rm eff}$.
At $T\sim 1$ MeV, before BBN begins, neutrinos have chemically decoupled from SM plasma and thus evolve adiabatically like dark radiation. We can therefore absorb the density of dark radiation, $\rho_{DR}$, into an extra neutrino component \cite{Aghanim:2018eyx},

\begin{align}\label{eq:deltaN_rho}
\Delta N_{\rm eff}
=\left[\frac{8}{7}\left(\frac{11}{4}\right)^{4/3}\frac{\rho_{DR}}{\rho_{\gamma}}\right]_{\rm today}.
\end{align}
Thus $\Delta N_{\rm eff}$ depends on the ratio of homogeneous densities of dark-radiation to SM. 

In appendix \ref{appendix:island_universe}, using the separate universe principle \cite{Weinberg:2003sw, Lyth:2003ip}, we show that long wavelength dark radiation isocurvature, $S_{\textrm{DR}}$, can  be absorbed into the homogeneous densities for super-horizon sized patches. Isocurvature between the dark-radiation and the photon bath is defined as
\begin{align}\label{eq:S_dr_def}
S_{\textrm{DR}}=\frac{3}{4}\Big(\frac{\delta\rho_{DR}}{{\rho}_{DR}}-\frac{\delta\rho_{\gamma}}{{\rho}_{\gamma}}\Big).
\end{align}
Since the photon fluctuations are assumed to be adiabatic together with the SM density fluctuations,  isocurvature  between the dark-radiation and the photon bath is equivalent to isocurvature between the dark-radiation and the SM plasma.
A super-horizon patch characterized by some window function, $W_{\lambda/2}$, that has support within a sphere of radius $\lambda/2$, has an average isocurvature, $\Delta S_{DR}$, given by
\begin{align}\label{eq:def_sdr}
\Delta S_{DR}=\frac{\int_0^{\infty} d^3x\, S_{DR}(\vec{x})W_{\lambda/2}(\vec{x})}{\int d^3x\, W_{\lambda/2}(\vec{x})}.
\end{align}
The dark-radiation isocurvature causes each such patch to observe $\DN$ given by (see appendix~\ref{sec:app})
\begin{align}\label{eq: deltaN_iso}
\Delta N_{\rm eff}= \Delta \bar{N}_{\textrm{eff}}\left(\frac{1+\frac{4}{3}\frac{1}{1+{\rho}_{DR}/{\rho}_{SM}}\Delta S_{DR}}{1-\frac{4}{3}\frac{{\rho}_{DR}/{\rho}_{SM}}{1+{\rho}_{DR}/{\rho}_{SM}}\Delta S_{DR}}\right) \approx \Delta \bar{N}_{\textrm{eff}}\left(1+\frac{4}{3}\Delta S_{DR}\right) ,
\end{align}
where the over-bar denotes a spatial average, and $\rho_{SM}$ is the density of SM plasma. In the second equality we have assumed that $\rho_{DR}\Delta S_{DR}/\rho_{SM}\ll1$, or equivalently $\Delta \bar{N}_{\textrm{eff}}\Delta S_{DR}\ll 1$. 
Because $\Delta S_{DR}$ takes different values in different regions, dark-radiation isocurvature leads to spatial variations in $\Delta N_{\rm eff}$.

In the presence of a dark-radiation isocurvature mode, regions of the Universe that were causally disconnected during BBN have different primordial abundances of light elements due to their different values of $\DN$. For example, the D/H ratio, $D$, is primarily a function of $\DN$ and the baryon fraction $\Omega_b h^2$. Assuming small fluctuations in $\DN$, the fluctuation in $D$ is given by
\begin{align}
    D& \approx \bar{D}+\frac{\partial D}{\partial \DN}\bigg|_{\Delta \bar{N}_{\rm eff}}(\DN-\Delta \bar{N}_{\rm eff}),\label{eq:D_fluc}
\end{align}
where $\bar{D}=D(\Delta \bar{N}_{\rm eff},\Omega_b h^2)$. This gives us a direct relation between the variance in $D$, given by $\sigma_d$, and the variance in $\DN$ fluctuations
\begin{align}\label{eq:sigma_D}
    \sigma_d=\frac{\partial D}{\partial \DN}\bigg|_{\Delta \bar{N}_{\rm eff}}\sigma_{N_{\textrm{eff}}}.
\end{align}
In practice, the derivatives, $\partial D/\partial \DN$, can be obtained numerically from publicly available codes. In this work, we use Parthenope \cite{Consiglio:2017pot}.

Immediately following BBN, the primordial abundances in the patches are conserved. As the Universe expands, and overdensities collapse to form galaxies, variations on scales smaller than those scales that collapse to form galaxies get mixed.\footnote{The ratio of BBN yields to Hydrogen can increase slightly during the collapse of structures \cite{medvigy2001element}. However, we ignore this effect as the increase is well below the current sensitivities of measurements. Moreover, a post-BBN diffusion of elements \cite{pospelov2012lithium} will erase differences in ${}^4$He/H or D/H ratios inside the diffusion volume. Our analysis is unaffected by this post-BBN diffusion as long as the diffusion length scales are smaller than the galactic-scales.} Measurements from different galaxies are therefore sensitive to isocurvature fluctuations down to galactic scales. Consequently, the scale $\lambda$ entering in eq.~\eqref{eq:def_sdr} is the comoving size of a patch, $\lambda_{\rm gal}$, which collapses to form the galaxies we observe
\begin{align}\label{eq:lambda_gal}
\lambda_{\rm gal}=3.7\left(\frac{M_{\rm gal}}{10^{12}M_{\odot}}\right)^{1/3}\left(\frac{\Omega_{\rm m}h^2}{0.14}\right)^{-1/3}\textrm{Mpc},
\end{align}
where $M_{\rm gal}$ is the mass of the galaxy. The scale $\lambda_{\rm gal}$ is larger than the horizon during BBN, $\sim $kpc, which implies that our analysis built on eq.~\eqref{eq: deltaN_iso} is self-consistent. 

The value of $\Delta N_{\rm eff}$ experienced by a galaxy is sampled from a distribution with mean $\Delta \bar{N}_{\rm eff}$ and variance $\sigma_{N_{\rm eff}}$. Moreover, $\sigma_{N_{\textrm{eff}}}$ is related to the power spectrum of isocurvature fluctuation, $P_S$, as
\begin{align}\label{eq:sigma_neff_iso2}
\sigma_{N_{\textrm{eff}}}^2=\frac{16}{9}\Delta \bar{N}_{\textrm{eff}}^2\langle \Delta S_{DR}^2\rangle=\frac{16}{9}\frac{\Delta \bar{N}_{\textrm{eff}}^2}{\left[\int d^3x\, W_{\lambda_{\rm gal}/2}(\vec{x})\right]^2}\int_0^{\infty} \frac{dk}{k} |W_{\lambda_{\rm gal}/2}(k)|^2\frac{k^3 P_{S}(k)}{2\pi^2}.
\end{align}
Since the details of galaxies providing ${}^4$He/H or D/H are usually not observable, the accurate estimation of $W_{\lambda_{\rm gal}/2}$ is not feasible. Consequently, we cannot exactly relate the variance in the average isocurvature experienced by a galaxy, $\langle \Delta S_{DR}^2\rangle$, to the isocurvature power spectrum. However, we can obtain an approximate relation between $\langle \Delta S_{DR}^2\rangle$ and $P_S$.
Assuming a blue-tilted isocurvature power spectrum
\begin{align}
\Delta_S^2\equiv \frac{k^3 P_S(k)}{(2\pi^2)} \propto  k^{n},
\end{align}
with $n>0$, and assuming a spherical Gaussian window function, $W_{\lambda/2}(k)=\exp(-k^2\lambda^2/8)$, eq.~\eqref{eq:sigma_neff_iso2} yields
\begin{align}\label{eq:sigma_neff_iso}
\langle \Delta S_{DR}^2\rangle= \frac{\Gamma(n/2)}{2}\Delta_S^2(2\lambda_{\rm gal}^{-1}).
\end{align}
Here $\Gamma(x)$ is the Euler Gamma function. As $\langle \Delta S_{DR}^2\rangle$ determines $\sigma_{N_{\textrm{eff}}}$ which in turn informs the variance in D/H (or ${}^4$He/H) ratios, the intrinsic variance in the D/H (or ${}^4$He/H) ratio in a given galaxy is determined by dark-radiation isocurvature at scales $\sim \lambda_{\rm gal}/2$.

\section{Constraints from ${}^4$He/H and D/H data}\label{sec:constraints}

In this section we use data from observations of the ratios of ${}^4$He/H and D/H to place constraints on dark-radiation isocurvature. We first describe the datasets which we use for our analysis and then describe our methodology for D/H and ${}^4$He/H data separately.

\subsection{Datasets}
D/H measurements are taken from gas clouds that are seen in absorption against the light of an unrelated background quasar \cite{1976A&A....50..461A}. Correspondingly, by looking at the frequency distribution of the light from the quasar, one can estimate the redshift of the gas cloud as well as the column densities of neutral Hydrogen and Deuterium atoms.

For our analysis we use the D/H measurements provided in Ref.\ \cite{Cooke:2017cwo}. The data uses measurements from seven damped Lyman-$\alpha$ systems\footnote{The damped Lyman-$\alpha$ systems are distinct from the Lyman-$\alpha$ forest systems which provide matter structure constraints around $\gtrsim 1$ Mpc. They are differentiated on the basis of the amount of neutral Hydrogen column densities, $N({\rm H\ I})$. Lyman-$\alpha$ forest systems are those with $N({\rm H\ I})<10^{17} \textrm{cm}^{-1}$ and damped Lyman-$\alpha$ systems are those with $2\times 10^{20} \textrm{cm}^{-1}<N({\rm H\ I})$ \cite{doi:10.1146/annurev.astro.42.053102.133950}.} around redshifts $z\sim 2-3$, that satisfy the strict selection criteria of precision stated in Ref.\ \cite{Cooke_2014}. To estimate the comoving scale in the early universe from which the gas cloud formed, we require the mass of the gas cloud. While the masses of individual damped Lyman-$\alpha$ systems are not known, their masses have been estimated to be in the range $10^{11}-10^{12} M_{\odot}$ \cite{2019MNRAS.tmp.1435M, 2012JCAP...11..059F}.

The $^4{}$He abundance is derived from observations of the helium and hydrogen emission lines from H II regions in low-metallicity blue compact dwarf galaxies that have undergone little chemical evolution \cite{1974ApJ...193..327P}. Regions with minimal chemical evolution are selected so as to minimize ${}^4$He enrichment by stellar processes. However, there still remains some contamination that leads to an increase in the ${}^4$He/H ratio over its primordial value.

In this study we use ${}^4$He/H data provided in Ref.\ \cite{Aver:2015iza}. The data consists of 15 measurements of He II regions from 14 different galaxies. For our analysis, we assume that each galaxy has a uniform value of the primordial ${}^4$He/H ratio. Correspondingly, we combine the two measurements of the same galaxy into a single data point using a weighted average. Unlike in the case of Deuterium measurements, the galaxies providing Helium measurements have low redshifts $z<0.05$. Out of the 14 galaxies used in measuring ${}^4$He/H abundance we find the masses of three\footnote{The galaxies of whose masses we found are aliased as Mrk 209, Mrk 71 and SBS 1415$+$437 in \cite{Aver:2015iza}. While their aliases used in SPARC database are UGCA 281, NGA 2366, and PGC51017, respectively.} of them in the SPARC database \cite{Li:2019zvm}. Their masses are in the range $10^{10.2}-10^{10.6} M_{\odot}$.

\subsection{Constraints from D/H data}

The gas in damped Lyman-$\alpha$ systems is assumed not to have produced or destroyed significant amounts of Deuterium. Correspondingly, the measurement from a gas cloud  samples the primordial value of $D$ which is assumed to be drawn from a distribution with mean and variance given by $\{\bar{D},\sigma_d\}$. The probability of getting a measurement of $D_i$ from gas cloud $i$ is then given by
\begin{align}
    P(D_i|\{\bar{D},\sigma_d\})=&\frac{1}{\sqrt{2\pi(\sigma_{n,i}^2+\sigma_d^2)}}\exp\left(-\frac{(D_i-\bar{D})^2}{2(\sigma_{n,i}^2+\sigma_d^2)}\right),
\end{align}
where $\sigma_{n,i}$ is the estimated noise in the measurement of $D$. We have assumed that each measurement has the same intrinsic variance in $D$. We do so because the damped Lyman-$\alpha$ systems typically have masses in the relatively narrow range $10^{11}-10^{12}M_{\odot}$ \cite{2019MNRAS.tmp.1435M, 2012JCAP...11..059F}. Correspondingly, the gas clouds in our data have roughly the same $\lambda_{\rm gal}$ (see eq.~\eqref{eq:lambda_gal}) and thus the same variance in $D$ (see eqs.~\eqref{eq:sigma_neff_iso} and \eqref{eq:sigma_D}). Moreover, we have neglected covariance between different measurements. This approximation is valid because isocurvature on the scales of separation between different galaxies in our data (usually $> 100$ Mpc) is constrained by CMB measurements \cite{Akrami:2018odb} to be much smaller than the variance to which our analysis is sensitive.

The constraints from $D_i$ measurements are degenerate in $\DN$ and $\Omega_b h^2$. To remove this degeneracy we fix the value of $\Omega_b h^2$ using the Planck data,\footnote{The Planck constraints on $\Omega_b h^2$ are slightly degenerate with $N_{\rm eff}$. Correspondingly we take Planck constraints on $\Omega_b h^2$ values marignalised over $N_{\rm eff}$ from TT+TE+EE+lowl+lowlE+BAO data.} $\Omega_b h^2=0.02239\pm0.00018\equiv \bar{\Omega}_b h^2\pm \sigma_{\Omega_b}$ \cite{Aghanim:2018eyx}, where $\sigma_{\Omega_b}$ is the uncertainty on the baryon density, which is assumed to be spatially homogeneous. The corresponding likelihood function is then given by
\begin{align}\label{eq:basic_L}
    \mathcal{L}_0(\DN,\sigma_{\Delta N_{\textrm{eff}}})=\int_{0}^{\infty}\left[\prod_i P(D_i|\{\bar{D},\sigma_d\})\right]_{\DN,\Omega_b h^2,\sigma_{N_{\textrm{eff}}}}\frac{\exp\Big(-\frac{(\Omega_b h^2-\bar{\Omega}_b h^2)^2}{2\sigma_{\Omega_b}^2}\Big)}{\sqrt{2\pi\sigma_{\Omega_b}^2}} d(\Omega_b h^2).
\end{align}
We numerically marginalize over $\Omega_b h^2$ to obtain our likelihood estimate.

\begin{figure}
\begin{subfigure}{.5\textwidth}
\includegraphics[width=1.00\textwidth]{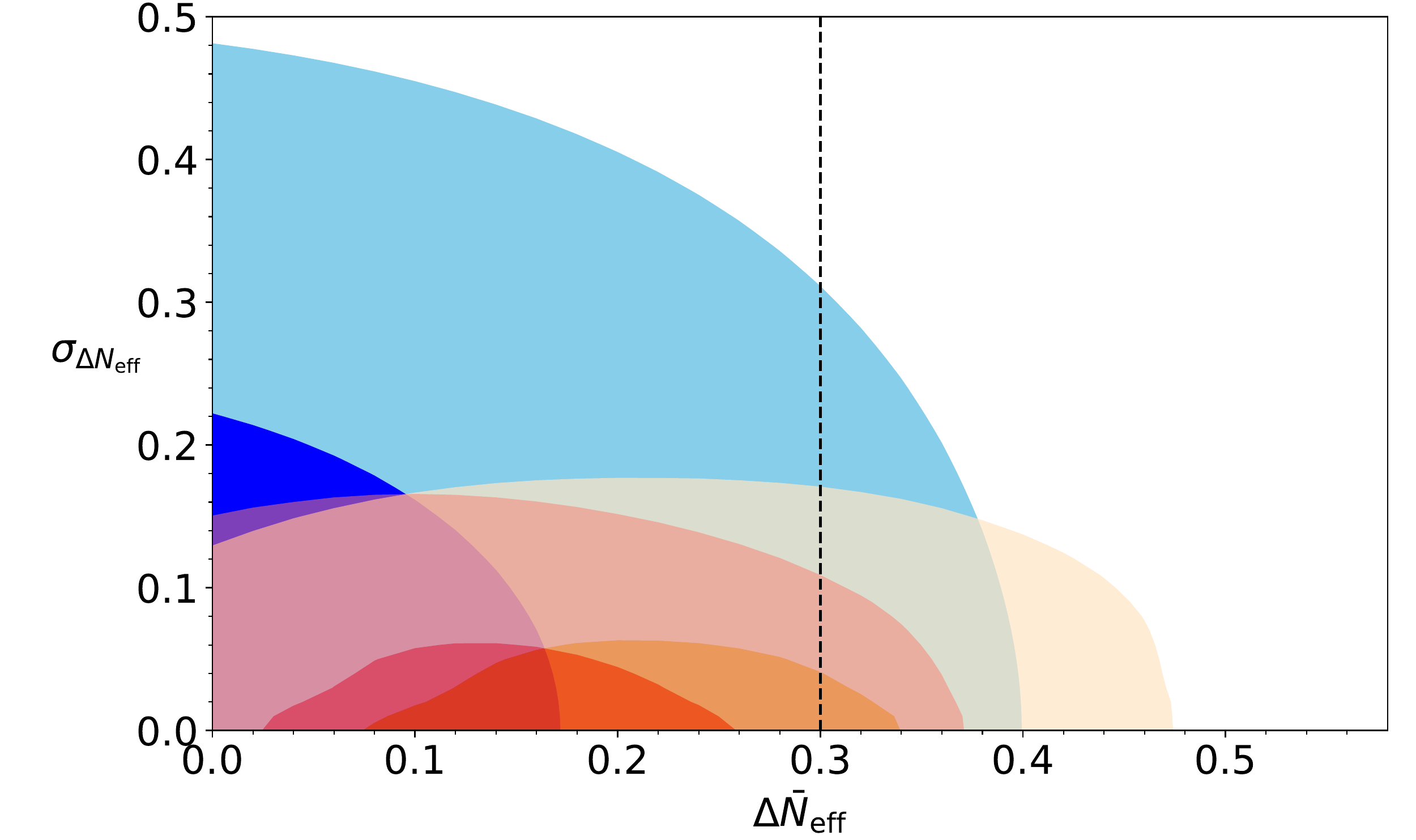}
\end{subfigure}
\begin{subfigure}{.5\textwidth}
\includegraphics[width=1.00\textwidth]{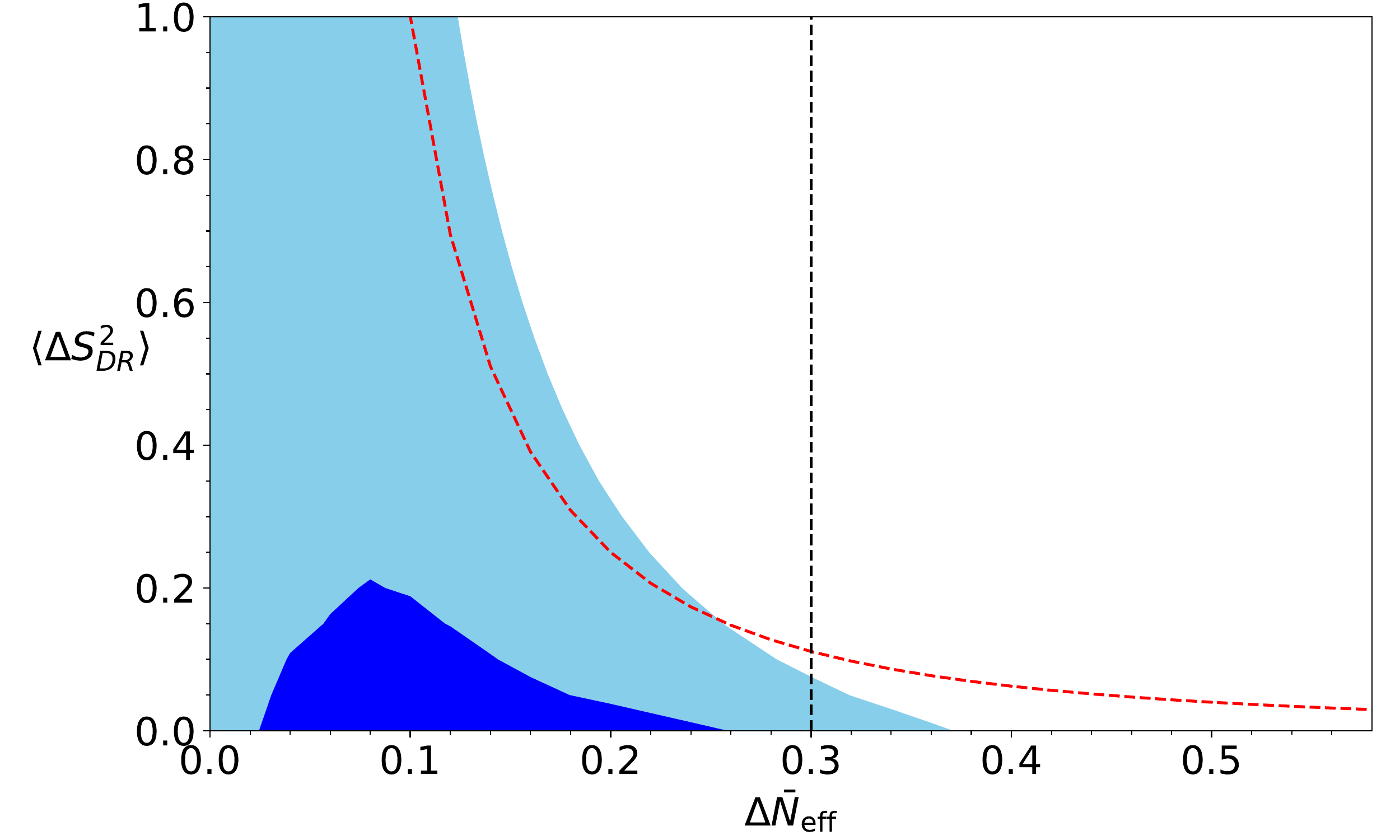}
\end{subfigure}
\caption{\textit{Left}: $1\sigma$ and $2\sigma$ constraints on $\{ \Delta \bar{N}_{\rm eff}, \sigma_{\Delta N_{\rm eff}}\}$ from ${}^4$He/H data (blue contours), D/H data (orange contours) and from combined data (red contours). The vertical dashed line denotes the upper limit on $\DN$ from Planck data with $95\%$ confidence \cite{Aghanim:2018eyx}. \textit{Right}: Constraints on $\{ \Delta \bar{N}_{\rm eff}, \langle \Delta S_{\textrm{DR}}^2 \rangle\}$ from combined data. Here $\Delta S_{\textrm{DR}}$ is the average dark radiation in a galaxy. For blue-tilted isocurvature mode $\langle \Delta S_{\textrm{DR}}^2\rangle$ is approximately same as the normalized isocurvature power spectrum at 1 Mpc, $\Delta_S^2(1 \textrm{Mpc}^{-1})$ (see  eq.~\eqref{eq:sigma_neff_iso}). The red-dashed line marks the parameters space at which $\Delta \bar{N}_{\rm eff}\Delta S_{\textrm{DR}}=0.1$. Correspondingly, the small $\Delta \bar{N}_{\rm eff}\Delta S_{\textrm{DR}}$ approximation made in eq.~\eqref{eq: deltaN_iso} holds for most of our parameter space.}
\label{fig:constraint}
\end{figure}

Using Parthenope \cite{Consiglio:2017pot} to estimate $D(\DN,\Omega_b h^2)$ and $\frac{\partial D}{\partial \DN}$, we find the 1$\sigma$ and $2\sigma$ limits on $\{ \Delta \bar{N}_{\rm eff}, \sigma_{N_{\textrm{eff}}}\}$ shown as orange contours in left panel of figure~\ref{fig:constraint}.

\subsection{Constraints from ${}^4$He/H data}
The methodology used in analysing D/H data is also applicable for ${}^4$He/H data after accounting for the ${}^4$He produced by stellar processes. To estimate the amount of primordial ${}^4$He/H ratio, $Y_p$, in a given galaxy we assume a linear relation between the Oxygen to Hydrogen ratio (O/H) and the ${}^4$He/H ratio produced by stellar processes. Thus the net ${}^4$He/H ratio, $Y$, found in a galaxy is given by
\begin{align}
Y=Y_p+m\times ({\rm O/H}),
\end{align}
where $m$ is the proportionality between O/H production and ${}^4$He/H production through stellar processes.
Above, $Y_p$ fluctuates with $\DN$ in a similar manner as in eq.~\eqref{eq:D_fluc}, except with $D$ replaced by $Y_p$. Similarly, the variance in $Y_p$, given by $\sigma_y$, and the variance in $\DN$ fluctuations are related by eq.~\eqref{eq:sigma_D}, except with $D$ replaced by $Y_p$.

Taking into account the linear relation between O/H and $Y$, the probability of getting a measurement of $Y_i$ from galaxy $i$ is given by
\begin{align}
    \tilde{P}(Y_i|\{\bar{Y}_p,\sigma_y,m\})=&\frac{1}{\sqrt{2\pi(\sigma_{n,i}^2+\sigma_y^2+m^2\sigma_{{\rm O}n,i}^2)}}\exp\Big(-\frac{(Y_i-(\bar{Y}_p+m({\rm O/H})_i))^2}{2(\sigma_{n,i}^2+\sigma_y^2+m^2\sigma_{{\rm O}n,i}^2)}\Big),
\end{align}
where $\sigma_{{\rm O}n,i}$ is the noise in O/H measurement. Just like in the case of Deuterium, we have considered all galaxies to have the same variance in $\DN$ and neglected covariance between different galaxies. We marginalize over $\Omega_bh^2$ as we did for Deuterium in eq.~\eqref{eq:basic_L}. Additionally, as the precise value of $m$ is unknown, we explicitly marginalize over $m$ assuming a uniform prior,
\begin{align}
    \mathcal{L}(\DN,\sigma_{N_{\textrm{eff}}})=&\int_{0}^{\infty}\int_0^{\infty} \left[\prod_i \tilde{P}(Y_i|\{\bar{Y},\sigma_y,m\})\right]_{\DN,\Omega_b h^2,\sigma_{N_{\textrm{eff}}}}\nonumber\\
    &\times\frac{\exp\Big(-\frac{(\Omega_b h^2-\bar{\Omega}_b h^2)^2}{2\sigma_{\Omega_b}^2}\Big)}{\sqrt{2\pi\sigma_{\Omega_b}^2}} dm\ d(\Omega_b h^2).
\end{align}
Since the ${}^4$He/H and O/H data prefers values of $m\sim O(10^2)$ \cite{Aver:2015iza}, we have $m\sigma_{{\rm O}n,i}\sim 10^{-2}\sigma_{n,i}$. Thus we neglect the contribution from $m^2\sigma_{{\rm O}n,i}^2$ terms in our likelihood function. Consequently, using the definition of $\mathcal{L}_0$ in eq.~\eqref{eq:basic_L} but with $D$ replaced by $Y$, the above integral simplifies to,
\begin{align}
    \mathcal{L}(\DN,\sigma_{N_{\textrm{eff}}})\approx&\mathcal{L}_0(\DN,\sigma_{N_{\textrm{eff}}}) \nonumber\\
    &\times \int_0^{\infty} \exp\left(-\frac{1}{2}\left[\sum_i\frac{({\rm O/H})_i^2}{(\sigma_{n,i}^2 +\sigma_y^2)}\right]m^2+\left[\sum_i\frac{(Y_i-\bar{Y})({\rm O/H})_i}{(\sigma_{n,i}^2 +\sigma_y^2)}\right]m \right) dm\nonumber\\
    =&\mathcal{L}_0(\DN,\sigma_{N_{\textrm{eff}}}) \frac{\sqrt{\pi} e^{b^{2} /(2 a)}\left(\operatorname{erf}\left(\frac{b}{\sqrt{2a}}\right)+1\right)}{\sqrt{a}},
\end{align}
where $a$ and $b$ are the inverse variance weighted sum of $({\rm O/H})_i^2$ and $(Y_i-\bar{Y})({\rm O/H})_i$ respectively (terms in the square bracket in the first line).
Using Parthenope \cite{Consiglio:2017pot} to estimate $Y(\DN,\Omega_b h^2)$ and $\frac{\partial Y}{\partial \DN}$ we find the 1$\sigma$ and $2\sigma$ limits on $\{ \Delta \bar{N}_{\rm eff}, \sigma_{N_{\textrm{eff}}} \}$, shown as blue contours in left panel of figure~\ref{fig:constraint}.

The red contours in the left panel of figure~\ref{fig:constraint} show the combined constraints from Helium and Deuterium data, which restricts the variance in $\DN$ to be $\sigma_{N_{\textrm{eff}}}\leq 0.17$ at 95\% confidence. In the right panel of figure~\ref{fig:constraint} we convert the constraints on $\sigma_{N_{\textrm{eff}}}$ to constraints on $\langle \Delta S_{\textrm{DR}}^2\rangle$ using eq.\ \eqref{eq:sigma_neff_iso2}. Since the masses of the galaxies used in our measurements lie in the range $10^{10}-10^{12}M_{\odot}$, we consider all our measurements to have $\lambda_{\rm gal}\sim 2$ Mpc (see eq.~\eqref{eq:lambda_gal}). Correspondingly, the variance in average dark radiation isocurvature in a galaxy, $\langle \Delta S_{\textrm{DR}}^2\rangle$ approximately measures isocurvature on scales around $\lambda_{\rm gal}/2\sim1$ Mpc, i.e. $\langle \Delta S_{\textrm{DR}}^2\rangle \sim \Delta_S^2(k \sim 1 \textrm{Mpc}^{-1})$ (see eq.~\eqref{eq:sigma_neff_iso}). The constraints on isocurvature become significantly weaker for smaller values of $\Delta\bar{N}_{\rm eff}$. This is indicative of the fact that smaller dark radiation densities make it harder for the isocurvature component to gravitationally affect BBN.

\section{Discussion and conclusions}\label{sec:conclusion}

In a universe with a dark radiation field that is populated independently of the SM sector  following inflation, an isocurvature mode can naturally occur between the two sectors. In this work, we have demonstrated that such an isocurvature mode leads to spatially varying BBN yields. Correspondingly, we have derived constraints on the existence of an isocurvature mode between SM plasma and putative dark radiation by looking at spatial variations in ${}^4$He/H and D/H abundances. 

A lack of excess variance in observed  ${}^4$He/H and D/H data  limits the amount of isocurvature present during BBN. Assuming each galaxy has internally uniform ${}^4$He/H and D/H ratios, a single galaxy probes dark radiation isocurvature at scales $\sim \lambda_{\rm gal}/2$. Here $\lambda_{\rm gal}$ corresponds to the comoving size of the overdensity which eventually collapses to form the galaxy in question. Since the structures which provide ${}^4$He/H or D/H measurements typically have masses around $10^{11}~M_{\odot}$, our analysis is sensitive to dark radiation isocurvature at scales $\sim 1$ Mpc. Subsequently, using ${}^4$He/H data from measurements of nearby galaxies \cite{Cooke:2017cwo} and D/H data from measurements of high-redshift Lyman-$\alpha$ absorption systems \cite{Aver:2015iza}, we constrained the variance of average dark radiation isocurvature fluctuations,  to be $\sqrt{\langle \Delta S_{\textrm{DR}}^2\rangle} <0.13/\Delta \bar{N}_{\rm eff}$ (see right panel of figure~\ref{fig:constraint}) at 95\% confidence. The quantity $\langle \Delta S_{\textrm{DR}}^2\rangle$ is approximately the same as the normalized isocurvature power spectrum at 1 Mpc. The exact relation between $\langle \Delta S_{\textrm{DR}}^2\rangle$ and the isocurvature power spectrum requires an accurate estimation of the primordial overdensities that collapse to form the galaxies in our data. Finally, we also constrain the variance in $\Delta N_{\rm eff}$ to be $\sigma_{N_{\rm eff}}<0.17$ at 95\% confidence.

By translating the neutrino isocurvature constraints by Planck \cite{Akrami:2018odb} to dark radiation isocurvature,\footnote{CMB measurements cannot distinguish between the effects from dark-radiation and neutrinos. Consequently the isocurvature, $\mathcal{S}$ constrained by CMB would have contributions from both dark-radiation and neutrinos,
$\mathcal{S}= \frac{3}{4}\left(\frac{\delta\rho_{DR}+\delta\rho_{\nu}}{\rho_{DR}+\rho_{\nu}}-\frac{\delta\rho_{\gamma}}{\rho_{\gamma}}\right)=\frac{\rho_{DR}}{\rho_{DR}+\rho_{\nu}}\mathcal{S}_{DR}+\frac{\rho_{\nu}}{\rho_{DR}+\rho_{\nu}}\mathcal{S}_{\nu}.$
The subscript $\nu$ refer to quantities for neutrinos.} we find that dark radiation isocurvature fluctuations are constrained to be less than $\sim 10^{-5}\times N_{\rm eff}/\Delta \bar{N}_{\rm eff}$ on scales larger than $\sim 10$~Mpc. Although the CMB is much more sensitive probe than the ${}^4$He/H and D/H data, the latter is able to probe isocurvature at scales that are inaccessible to CMB measurements.

If dark radiation and dark matter fluctuations are correlated---which could occur, for example, in theories of dark freeze-out in the presence of isocurvature---then constraints from measurements of clustering in the Lyman-$\alpha$ forest on dark matter isocurvature can be translated to dark radiation. The Lyman-$\alpha$ forest data constraints DM isocurvature to be less than $10^{-4}$ at $1$ Mpc \cite{Beltran:2005gr}, by putting limits on the excess power over the adiabatic matter power spectrum extrapolated from CMB measurements. In contrast, the analysis in this study is sensitive directly to the isocurvature mode between dark radiation and the SM sector, and is unaffected by the adiabatic fluctuations.

The extension of constraints from Lyman-$\alpha$ forest data to smaller scales is limited by solving non-linear structure formation---a complication that does not affect our analysis. In contrast, the techniques used in this study can theoretically be used to extend the constraints down to scales $\sim 0.01$ Mpc, i.e.\ scales slightly larger than the comoving horizon during BBN. To achieve the constraints at such small scales one would require measurements of ${}^4$He/H or D/H from structures with masses of order $\sim 10^6-10^7 M_{\odot}$. Potential future measurements of ${}^4$He/H  in halos of masses $\sim 10^{9.5} M_{\odot}$ \cite{Sykes:2019sop}  would extend the constraints down to $\sim 0.3$ Mpc.

\acknowledgments

We thank Brian Fields for useful discussions. We thank Ofelia Pisanti and Ken Nollett for providing help in implementing Parthenope. We thank Robert Scherrer for discussion regarding post-BBN diffusion. The work of PA and PR is supported in part by NASA Astrophysics Theory Grant NNX17AG48G. PA thanks  the Kavli Institute for Theoretical Physics for hospitality and support through National Science Foundation Grant No.\ NSF PHY-1748958 as this work was nearing completion. This work was supported by Brand and Monica Fortner.

\appendix

\section{Dark radiation isocurvature as spatially varying $N_{\rm eff}$}
\label{appendix:island_universe}

In this appendix, we demonstrate that, on super-horizon scales, the effect of an isocurvature mode between dark radiation and the visible sector is to vary the local value of the effective number of degrees of freedom, $N_{\rm eff}$. We begin by reviewing the separate Universe principle, before showing explicitly how a dark-radiation isocurvature mode may be interpreted as a spatially varying effective number of relativistic species, $N_{\rm eff}(\vec{x})$.

\subsection{Separate universe principle and total density fluctuations} 

The separate universe principle  (see, e.g., Refs.\ \cite{Weinberg:2003sw, Lyth:2003ip}) posits that each super-horizon sized patch can be treated as an isolated, independent,  Friedman-Robertson-Walker (FRW) universe. Any fluctuations on scales larger than the horizon simply become a part of the background variables of that island universe.

We begin by demonstrating that zero-mode fluctuations (Fourier mode $k=0$) can be absorbed into the background quantities (density, pressure, etc.). The reason is straightforward. After fixing our coordinate system (fixing the gauge), there remains a residual coordinate transformation which can be used to absorb the fluctuations into a redefinition of the background. This can be seen explicitly as follows. We write the perturbed FRW metric as
\begin{align}\label{eqn:genpert}
    [g_{\mu\nu}]=\begin{bmatrix}
-(1+2\psi) & -a\partial_i\beta\\
-a\partial_i\beta & a^2[(1+2\phi)\delta_{ij}+2\partial_i\partial_j\gamma]\\
\end{bmatrix}.
\end{align}
Under an infinitesimal coordinate transformation (gauge transformation) $x^{\mu}{}'=x^{\mu}+\xi^{\mu}$, with $\xi_{i}=\partial_i\xi$, the scalar parts of the metric perturbations transform as
\begin{align}\nn
    \psi'=\psi-\dot{\xi}^0,&&
    \beta'=\beta-\frac{1}{a}\xi^0+a\dot{\xi},\\
    \phi'=\phi-H\xi^0,&&
    \gamma'=\gamma-\xi.
\end{align}
The two scalar parts of the coordinate transformation, $\xi^0$ and $\xi$, can be used to set two of the four scalar perturbations in eq.\ \eqref{eqn:genpert} to zero. In conformal Newton gauge,  $\xi^0$ and $\xi$ are chosen to  set $\beta=\gamma=0$ and make  the metric diagonal. However, because only gradients of $\beta$ and $\gamma$ appear in the metric, spatially uniform changes of coordinates leave the metric diagonal. Specifically, consider a transformation of the conformal Newton metric given by
\begin{align}\label{eq:residual_gauge2}
      \xi^0=\epsilon(t)&&
      \xi=\omega x^ix^j\delta_{ij},
\end{align}
where $\omega$ is a constant. The metric perturbations transform as
\begin{align}
      \psi'=\psi-\dot{\epsilon}&&
    \beta'=-\frac{1}{a}\epsilon(t),\\
    \phi'=\phi-H\epsilon(t)&&
    \gamma'=-\omega x^ix^j\delta_{ij}.
\end{align}
Because only spatial derivatives of $\beta$ and $\gamma$ appear in the metric, the transformation in eq.\ \eqref{eq:residual_gauge2} leaves the metric diagonal. The diagonal term from $\gamma'$ can be absorbed into $\phi$ to give
\begin{flalign}
      \psi'&=\psi-\dot{\epsilon},\label{eq:residual_psi}\\
      \phi'&=\phi-H\epsilon(t)+\partial_i\partial_i\gamma'/3=\phi-H\epsilon(t)-2\omega.\label{eq:residual_phi}
\end{flalign}
The transformation as described in eq.~\eqref{eq:residual_gauge2} keeps the metric diagonal and is a residual gauge freedom for conformal Newton gauge. This residual gauge freedom in eqs.~\eqref{eq:residual_psi} and \eqref{eq:residual_phi} can be used to remove the spatially homogeneous fluctuations in $\phi$ and $\psi$, and set the $k = 0$ fourier mode to zero, $\phi'_{k = 0} = \psi'_{k = 0} = 0$, giving the relations
\begin{align}\label{eq:residual_gauge_to_flat}
       H\epsilon+2\omega=\phi_{k=0},&&
       \dot{\epsilon}=\psi_{k=0}.
\end{align}

In the absence of metric perturbations, the Einstein equations imply that the total density perturbation vanishes. We can demonstrate this explicitly by noting that, under the residual gauge transformation, eq.~\eqref{eq:residual_gauge2}, the density perturbation transforms as
\begin{align}\label{eq:densitytransform}
    \delta\rho'&=\delta\rho-\dot{\rho}\epsilon=\frac{1}{4\pi G}(4\pi G\delta\rho-3H\dot{H}\epsilon),
\end{align}
where in the second equality we used Friedmann equation for the background, $H^2=8\pi G\rho/3$. Further, because $\omega = {\rm const.}$, eq.~\eqref{eq:residual_gauge_to_flat} gives us the relation
\begin{align}
    \dot{H}\epsilon=\dot{\phi}_{k=0}-H\psi_{k=0}.
\end{align}
Substituting this result into the gauge transformation given in eq.\ \eqref{eq:densitytransform} for $k=0$ mode, we obtain
\begin{align}
     \delta\rho'_{k=0}&=\frac{1}{4\pi G}(4\pi G\delta\rho_{k=0}-3H\dot{\phi}_{k=0}+3H^2\psi_{k=0}).
\end{align}
The above vanishes identically after using the Einstein equation for $\delta\rho_{k=0}$.  The residual gauge transformation in eq.~\eqref{eq:residual_gauge2}  self-consistently removes all zero mode perturbations in the total density. The shift $\epsilon$  that is required to gauge away the perturbations is simply a uniform shift in coordinate time. 

While the above analysis holds exactly for $k=0$ mode, it  can be extended to superhorizon modes up to corrections of order $O(k^2/(aH)^2)$.  To see this, consider a Universe with a single superhorizon mode fluctuation, $\phi_k$ with $k \ll a H$. On a patch of the Universe with a scale sufficiently small compared to $k^{-1}$, but still large compared to the horizon, the fluctuation $\phi_k$ appears almost constant. If we consider two such patches separated by a distance comparable to $k^{-1}$, each patch samples a different approximately uniform value of $\phi_k$. Consequently, in each patch, we may use the residual gauge freedom to remove this approximately constant $\phi_k$. The approximately uniform shift in coordinate needed to make each patch uniform is different in each patch. Due to the equivalence principle, the only observable effects of such a shift enter at order $O(k^2/(aH)^2)$.

\subsection{Adiabatic vs isocurvature perturbations and residual gauge shifts}\label{sec:iso_with_gauge}
Above we showed that the total density fluctuations can be removed on super-horizon scales. However, if the Universe is filled with a multi-component fluid, the perturbations in each fluid species need not necessarily vanish after this procedure. In fact, only adiabatic perturbations in each species necessarily vanish, as we now demonstrate. Adiabatic perturbations between different species are related by
\begin{align}\label{eq:adiabatic_def}
\frac{\delta\rho_1}{\dot{\rho}_1}=\frac{\delta\rho_2}{\dot{\rho}_2}=\frac{\delta\rho_3}{\dot{\rho}_3}=\ldots =\epsilon
\end{align}
where $\epsilon$ is the quantity obtained by solving eq.~\eqref{eq:residual_gauge_to_flat}. We also show that isocurvature perturbations do not in general vanish, but can be absorbed into the background quantites within each patch. In a Universe with isocurvature perturbations, separate Universes evolve with both shifted clocks due to the background adiabatic fluctuations, but also spatially varying background densities, as we now demonstrate.

Consider now a universe comprised of two non-interacting perfect fluids, with densities $\rho_1$ and $\rho_2$. We suppose that on large scales $k^{-1} \gg (aH)^{-1}$, there exists density fluctuations in both species. We then consider a patch of the Universe small compared to $k^{-1}$ but large compared to the horizon.
In this patch, after the residual gauge shift with magnitude specified by eq. \eqref{eq:residual_gauge_to_flat}, the total density perturbation in this patch vanishes giving
\begin{flalign}
       \delta\rho'_{tot}=\delta\rho_1'+\delta\rho_2'=0.
\end{flalign}
However, note that this only constrains the total density to vanish; it is not necessary for individual $\delta\rho_i'$ to also vanish. When they do, then using eq.~\eqref{eq:densitytransform}, we see that the species satisfy eq.~\eqref{eq:adiabatic_def}.

For an isocurvature mode, the density perturbations in individual species do not vanish after the gauge shift
\begin{flalign}\label{eq:iso_12}
       \delta\rho_1'=-\delta\rho_2'\neq 0.
\end{flalign}
However, these isocurvature perturbations can be absorbed into the background variables. To see this explicitly we consider the zero-mode density-perturbation equation,
\begin{flalign}
       \delta\dot{\rho}_i+3H(1+w_i)\delta \rho_i + 3\rho_i\dot{\phi}&=0,
\end{flalign}
where $i=1,2$ and $w_i$ is the equation of state of the $i$-th perfect fluid. After gauge shift this becomes
\begin{flalign}
              \delta\dot{\rho}'_i+3H(1+w_i)\delta \rho_i'&=0,
\end{flalign}
which can be trivially absorbed into the a redefinition of the background density
\begin{align}\label{eq:rho_redefinition}
\tilde{\rho}_i=\rho_i+\delta\rho_i'.
\end{align} 
This redefinition will not affect the metric or other Einstein variables as they only depend on total density in the universe, which remains unchanged after the absorption of isocurvature perturbations.

\subsection{Dark-radiation isocurvature and the spatial variation of $N_{\rm eff}$}\label{sec:app}
In the main body of the paper we are primarily concerned with patches of size $\lambda_{\rm gal}\sim 1$ Mpc, the matter inside of which collapses to form the galaxies from which we obtain ${}^4$He/H and D/H measurements. Since the horizon size during BBN, $\sim 1 $~Kpc, is much smaller than $\lambda_{\rm gal}$, we can use the separate universe principle to calculate the variation in $\Delta N_{\rm eff}$ due to isocurvature between the dark radiation and the SM radiation bath in different galaxy-sized patches.

Consider a spherical volume of radius $r=\lambda_{\rm gal}/2$, the matter within which later collapses to form a galaxy. Let $\rho_1$ and $\rho_2$ be the homogeneous densities of the SM and dark radiation respectively. In the previous sub-section we showed that an isocurvature fluctuation, $\delta\rho'_i(k)$, of a superhorizon-sized mode can be absorbed into the homogeneous density. For the spherical patch we are considering, using the separate universe principle, we can absorb the net $\delta\rho'_i(\vec{x})$ inside the volume into the homogeneous density (using eq.~\eqref{eq:rho_redefinition} and eq.~\eqref{eq:iso_12}),
\begin{align}\label{eq:absorb_vol}
\tilde{\rho}_1=\rho_1+ \Delta\rho_2' && \tilde{\rho}_2=\rho_2-\Delta\rho_2',
\end{align}
where $\Delta\rho_2'=-\Delta\rho_1'$ is the average isocurvature fluctuation in the dark radiation inside a spherical volume of radius $r$,
\begin{align}\label{eq:avg_iso}
\Delta\rho_2'=\int_0^{\infty}\frac{d^3x}{V_r}\delta\rho_2'(\vec{x})W_r(\vec{x}).
\end{align}
Here $W_r$ is a window function which weights the integral to be within $r$ radius from origin and $V_r$ is the volume swept by the window function, $V_r=\int d^3xW_r(\vec{x})$. The spherical volume effectively has $\tilde{\rho}_i$ as its homogeneous density. Note that while $\Delta\rho_2'$ involves contributions from all Fourier modes, the contribution from modes $k^{-1}\ll r$ is suppressed. The suppression is because the small wavelength modes have around the same number of over-densities and under-densities in a patch much larger than the mode's wavelength. Consequently, the super-horizon assumption in eq.~\eqref{eq:absorb_vol} approximately holds as long as $r$ is super-horizon sized.

Due to the presence of isocurvature, $N_{\rm eff}$ inside the spherical volume is also modified
\begin{align}\label{eq:Dneff_temp}
\Delta N_{\rm eff}\propto \frac{\tilde{\rho}_2}{\tilde{\rho}_1}=\frac{\rho_2+\Delta\rho_2'}{\rho_1-\Delta\rho_2'}.
\end{align}
In the main body of paper we define dark radiation isocurvature with respect to photons in eq.~\eqref{eq:S_dr_def}. However, as photons are adiabatic with SM plasma the isocurvature peruturbation can equivalently be written as
\begin{align}
S_{DR}=\frac{3}{4}\left(\frac{\delta\rho_2}{\rho_2}-\frac{\delta\rho_1}{\rho_1}\right)=\frac{3}{4}\frac{\rho_1+\rho_2}{\rho_1\rho_2}\delta\rho_2'.
\end{align}
In the second equality above we have expressed $S_{DR}$ in the uniform density gauge where $\delta\rho_2'=-\delta\rho_1'$. Consequently, the average isocurvature in the spherical volume is given by
\begin{align}
\Delta S_{DR}=\int_0^{\infty}\frac{d^3x}{V_r}S_{DR}(\vec{x})W_r(\vec{x})=\frac{3}{4}\frac{\rho_1+\rho_2}{\rho_1\rho_2}\Delta\rho_2'.
\end{align}
Replacing above back in eq.~\eqref{eq:Dneff_temp}, we can rewrite $\Delta N_{\rm eff}$ as
\begin{align}\label{eq:app_DN_Sdr}
\Delta N_{\rm eff}\propto \frac{\rho_2}{\rho_1}\frac{1+\frac{4}{3}\frac{1}{1+\rho_2/\rho_1}\Delta S_{DR}}{1-\frac{4}{3}\frac{\rho_2/\rho_1}{1+\rho_2/\rho_1}\Delta S_{DR}}\xrightarrow{(\rho_2/\rho_1)\Delta S_{DR}\ll 1}\frac{\rho_2}{\rho_1}\left(1+\frac{4}{3}\Delta S_{DR}\right).
\end{align}
Thus $\Delta N_{\rm eff}$ is sensitive to the average isocurvature in the spherical volume within the window function.

While we have derived Eq.~\eqref{eq:app_DN_Sdr} in  uniform-density slicing, this equation is gauge invariant. Uniform density slicing makes transparent the relation between $\Delta N_{\rm eff}$ in a super-horizon patch and $\mathcal{S}_{DR}$, both of which are gauge-invariant. Although the Hubble rate has the same value everywhere in uniform density slicing, the presence of dark-radiation isocurvature causes the individual SM and dark radiation densities to be inhomogeneous. Consequently, in this slicing, the photon temperature is inhomogeneous. Because the photon temperature is the relevant clock during BBN, the effect of dark radiation isocurvature on BBN is most easily understood in the slicing where the temperature is uniform. In uniform SM temperature (density) slicing, the presence of dark radiation isocurvature causes the Hubble rate to be inhomogeneous.

Since $\Delta S_{DR}$ depends on position, it varies between galaxies. The variance in $\Delta S_{DR}$ is given by
\begin{align}
\langle \Delta S_{DR}^2\rangle= \frac{1}{V_r^2}\int \frac{dk}{k}|W_{r}(k)|^2\Delta^2_S(k),
\end{align}
where $\Delta_S^2=k^3P_{S}(k)/(2\pi^2)$, and $W_{r}(k)$ and $P_{S}(k)$ are the Fourier transforms of $W_r(\vec{x})$ and $\langle S_{DR}(\vec{x})S_{DR}(\vec{x}\,{}') \rangle$ respectively. Since the contribution of small wavelength modes, $kr\gg 1$, is suppressed in $r$ sized patch, $W_r(k)$ is very small for $k\gg r^{-1}$. Consequently for a blue-tilted isocurvature power spectrum ($\Delta^2_S \sim k^n$, with $n>0$), we have
\begin{align}
\langle \Delta S_{DR}^2\rangle\sim \Delta^2_S(k\sim r^{-1}).
\end{align}
Thus, isocurvature  on scales near $\lambda_{\rm gal}$ leads to fluctuations in primordial $\Delta N_{\rm eff}$ between different galaxies.

\bibliographystyle{utphys}
\bibliography{references}

\providecommand{\href}[2]{#2}\begingroup\raggedright\begin{thebibliography}{10}

\bibitem{Abazajian:2016yjj}
{\bfseries CMB-S4} Collaboration, K.~N. Abazajian {\em et~al.}, ``{CMB-S4
  Science Book, First Edition},''
\href{http://arxiv.org/abs/1610.02743}{{\ttfamily arXiv:1610.02743
  [astro-ph.CO]}}.

\bibitem{Mangano_2005}
G.~Mangano, G.~Miele, S.~Pastor, T.~Pinto, O.~Pisanti, and P.~D. Serpico,
  ``Relic neutrino decoupling including flavour oscillations,''
  \href{http://dx.doi.org/10.1016/j.nuclphysb.2005.09.041}{{\em Nuclear Physics
  B} {\bfseries 729} no.~1-2, (Nov, 2005) 221–234}.
  \url{http://dx.doi.org/10.1016/j.nuclphysb.2005.09.041}.

\bibitem{Grohs_2016}
E.~Grohs, G.~Fuller, C.~Kishimoto, M.~Paris, and A.~Vlasenko, ``Neutrino energy
  transport in weak decoupling and big bang nucleosynthesis,''
  \href{http://dx.doi.org/10.1103/physrevd.93.083522}{{\em Physical Review D}
  {\bfseries 93} no.~8, (Apr, 2016) }.
  \url{http://dx.doi.org/10.1103/PhysRevD.93.083522}.

\bibitem{deSalas:2016ztq}
P.~F. de~Salas and S.~Pastor, ``{Relic neutrino decoupling with flavour
  oscillations revisited},''
  \href{http://dx.doi.org/10.1088/1475-7516/2016/07/051}{{\em JCAP} {\bfseries
  1607} (2016) 051},
\href{http://arxiv.org/abs/1606.06986}{{\ttfamily arXiv:1606.06986 [hep-ph]}}.

\bibitem{Akita:2020szl}
K.~Akita and M.~Yamaguchi, ``{A precision calculation of relic neutrino
  decoupling},'' \href{http://arxiv.org/abs/2005.07047}{{\ttfamily
  arXiv:2005.07047 [hep-ph]}}.

\bibitem{Abenza_2020}
M.~E. Abenza, ``Precision early universe thermodynamics made simple: Neff and
  neutrino decoupling in the standard model and beyond,''
  \href{http://dx.doi.org/10.1088/1475-7516/2020/05/048}{{\em Journal of
  Cosmology and Astroparticle Physics} {\bfseries 2020} no.~05, (May, 2020)
  048–048}. \url{http://dx.doi.org/10.1088/1475-7516/2020/05/048}.

\bibitem{Errard:2015cxa}
J.~Errard, S.~M. Feeney, H.~V. Peiris, and A.~H. Jaffe, ``{Robust forecasts on
  fundamental physics from the foreground-obscured, gravitationally-lensed CMB
  polarization},'' \href{http://dx.doi.org/10.1088/1475-7516/2016/03/052}{{\em
  JCAP} {\bfseries 1603} (2016) 052},
\href{http://arxiv.org/abs/1509.06770}{{\ttfamily arXiv:1509.06770
  [astro-ph.CO]}}.

\bibitem{Weinberg:2004kf}
S.~Weinberg, ``{Must cosmological perturbations remain non-adiabatic after
  multi-field inflation?},''
  \href{http://dx.doi.org/10.1103/PhysRevD.70.083522}{{\em Phys. Rev.}
  {\bfseries D70} (2004) 083522},
\href{http://arxiv.org/abs/astro-ph/0405397}{{\ttfamily arXiv:astro-ph/0405397
  [astro-ph]}}.

\bibitem{Lyth:2002my}
D.~H. Lyth, C.~Ungarelli, and D.~Wands, ``{The Primordial density perturbation
  in the curvaton scenario},''
  \href{http://dx.doi.org/10.1103/PhysRevD.67.023503}{{\em Phys. Rev.}
  {\bfseries D67} (2003) 023503},
\href{http://arxiv.org/abs/astro-ph/0208055}{{\ttfamily arXiv:astro-ph/0208055
  [astro-ph]}}.

\bibitem{Kawasaki:2011rc}
M.~Kawasaki, K.~Miyamoto, K.~Nakayama, and T.~Sekiguchi, ``{Isocurvature
  perturbations in extra radiation},''
  \href{http://dx.doi.org/10.1088/1475-7516/2012/02/022}{{\em JCAP} {\bfseries
  1202} (2012) 022},
\href{http://arxiv.org/abs/1107.4962}{{\ttfamily arXiv:1107.4962
  [astro-ph.CO]}}.

\bibitem{Kawakami:2012ke}
E.~Kawakami, M.~Kawasaki, K.~Miyamoto, K.~Nakayama, and T.~Sekiguchi,
  ``{Non-Gaussian isocurvature perturbations in dark radiation},''
  \href{http://dx.doi.org/10.1088/1475-7516/2012/07/037}{{\em JCAP} {\bfseries
  1207} (2012) 037},
\href{http://arxiv.org/abs/1202.4890}{{\ttfamily arXiv:1202.4890
  [astro-ph.CO]}}.

\bibitem{Akrami:2018odb}
{\bfseries Planck} Collaboration, Y.~Akrami {\em et~al.}, ``{Planck 2018
  results. X. Constraints on inflation},''
\href{http://arxiv.org/abs/1807.06211}{{\ttfamily arXiv:1807.06211
  [astro-ph.CO]}}.

\bibitem{Akrami:2018vks}
{\bfseries Planck} Collaboration, Y.~Akrami {\em et~al.}, ``{Planck 2018
  results. I. Overview and the cosmological legacy of Planck},''
\href{http://arxiv.org/abs/1807.06205}{{\ttfamily arXiv:1807.06205
  [astro-ph.CO]}}.

\bibitem{2010ApJ...716..907H}
G.~P. {Holder}, K.~M. {Nollett}, and A.~{van Engelen}, ``{On Possible Variation
  in the Cosmological Baryon Fraction},''
  \href{http://arxiv.org/abs/0907.3919}{{\ttfamily arXiv:0907.3919
  [astro-ph.CO]}}.

\bibitem{Inomata:2018htm}
K.~Inomata, M.~Kawasaki, A.~Kusenko, and L.~Yang, ``{Big Bang Nucleosynthesis
  Constraint on Baryonic Isocurvature Perturbations},''
  \href{http://dx.doi.org/10.1088/1475-7516/2018/12/003}{{\em JCAP} {\bfseries
  1812} no.~12, (2018) 003},
\href{http://arxiv.org/abs/1806.00123}{{\ttfamily arXiv:1806.00123
  [astro-ph.CO]}}.

\bibitem{Copi:1996td}
C.~J. Copi, K.~A. Olive, and D.~N. Schramm, ``{Implications of a primordial
  origin for the dispersion in d/h in quasar absorption systems},'' {\em
  Submitted to: Astrophys. J.} (1996) ,
\href{http://arxiv.org/abs/astro-ph/9606156}{{\ttfamily arXiv:astro-ph/9606156
  [astro-ph]}}.

\bibitem{1973ApJ...179..343W}
R.~V. {Wagoner}, ``{Big-Bang Nucleosynthesis Revisited},''
  \href{http://dx.doi.org/10.1086/151873}{{\em ApJ} {\bfseries 179} (Jan, 1973)
  343--360}.

\bibitem{Barrow_2018}
J.~D. Barrow and R.~J. Scherrer, ``Constraining density fluctuations with big
  bang nucleosynthesis in the era of precision cosmology,''
  \href{http://dx.doi.org/10.1103/physrevd.98.043534}{{\em Physical Review D}
  {\bfseries 98} no.~4, (Aug, 2018) }.
  \url{http://dx.doi.org/10.1103/PhysRevD.98.043534}.

\bibitem{Cooke:2017cwo}
R.~J. Cooke, M.~Pettini, and C.~C. Steidel, ``{One Percent Determination of the
  Primordial Deuterium Abundance},''
  \href{http://dx.doi.org/10.3847/1538-4357/aaab53}{{\em Astrophys. J.}
  {\bfseries 855} no.~2, (2018) 102},
\href{http://arxiv.org/abs/1710.11129}{{\ttfamily arXiv:1710.11129
  [astro-ph.CO]}}.

\bibitem{Aver:2015iza}
E.~Aver, K.~A. Olive, and E.~D. Skillman, ``{The effects of He I $\lambda$10830
  on helium abundance determinations},''
  \href{http://dx.doi.org/10.1088/1475-7516/2015/07/011}{{\em JCAP} {\bfseries
  1507} no.~07, (2015) 011},
\href{http://arxiv.org/abs/1503.08146}{{\ttfamily arXiv:1503.08146
  [astro-ph.CO]}}.

\bibitem{Aghanim:2018eyx}
{\bfseries Planck} Collaboration, N.~Aghanim {\em et~al.}, ``{Planck 2018
  results. VI. Cosmological parameters},''
\href{http://arxiv.org/abs/1807.06209}{{\ttfamily arXiv:1807.06209
  [astro-ph.CO]}}.

\bibitem{Weinberg:2003sw}
S.~Weinberg, ``{Adiabatic modes in cosmology},''
  \href{http://dx.doi.org/10.1103/PhysRevD.67.123504}{{\em Phys. Rev.}
  {\bfseries D67} (2003) 123504},
\href{http://arxiv.org/abs/astro-ph/0302326}{{\ttfamily arXiv:astro-ph/0302326
  [astro-ph]}}.

\bibitem{Lyth:2003ip}
D.~H. Lyth and D.~Wands, ``{The CDM isocurvature perturbation in the curvaton
  scenario},'' \href{http://dx.doi.org/10.1103/PhysRevD.68.103516}{{\em Phys.
  Rev.} {\bfseries D68} (2003) 103516},
\href{http://arxiv.org/abs/astro-ph/0306500}{{\ttfamily arXiv:astro-ph/0306500
  [astro-ph]}}.

\bibitem{Consiglio:2017pot}
R.~Consiglio, P.~F. de~Salas, G.~Mangano, G.~Miele, S.~Pastor, and O.~Pisanti,
  ``{PArthENoPE reloaded},''
  \href{http://dx.doi.org/10.1016/j.cpc.2018.06.022}{{\em Comput. Phys.
  Commun.} {\bfseries 233} (2018) 237--242},
\href{http://arxiv.org/abs/1712.04378}{{\ttfamily arXiv:1712.04378
  [astro-ph.CO]}}.

\bibitem{medvigy2001element}
D.~Medvigy and A.~Loeb, ``Element diffusion during cosmological structure
  formation,'' 2001.

\bibitem{pospelov2012lithium}
M.~Pospelov and N.~Afshordi, ``Lithium diffusion in the post-recombination
  universe and spatial variation of [li/h],'' 2012.

\bibitem{1976A&A....50..461A}
T.~F. {Adams}, ``{The detectability of deuterium Lyman alpha in QSOs.},'' {\em
  A\& A} {\bfseries 50} (Aug., 1976) 461--462.

\bibitem{doi:10.1146/annurev.astro.42.053102.133950}
A.~M. Wolfe, E.~Gawiser, and J.~X. Prochaska, ``Damped ly$\alpha$ systems,''
  \href{http://dx.doi.org/10.1146/annurev.astro.42.053102.133950}{{\em Annual
  Review of Astronomy and Astrophysics} {\bfseries 43} no.~1, (2005) 861--918}.
  \url{https://doi.org/10.1146/annurev.astro.42.053102.133950}.

\bibitem{Cooke_2014}
R.~J. Cooke, M.~Pettini, R.~A. Jorgenson, M.~T. Murphy, and C.~C. Steidel,
  ``{PRECISION} {MEASURES} {OF} {THE} {PRIMORDIAL} {ABUNDANCE} {OF}
  {DEUTERIUM},'' \href{http://dx.doi.org/10.1088/0004-637x/781/1/31}{{\em The
  Astrophysical Journal} {\bfseries 781} no.~1, (Jan, 2014) 31}.
  \url{https://doi.org/10.1088%2F0004-637x%2F781%2F1%2F31}.

\bibitem{2019MNRAS.tmp.1435M}
R.~{Mackenzie}, M.~{Fumagalli}, T.~{Theuns}, D.~J. {Hatton}, T.~{Garel},
  S.~{Cantalupo}, L.~{Christensen}, J.~P.~U. {Fynbo}, N.~{Kanekar},
  P.~{M{\o}ller}, J.~{O'Meara}, J.~X. {Prochaska}, M.~{Rafelski}, T.~{Shanks},
  and J.~{Trayford}, ``{Linking gas and galaxies at high redshift: MUSE surveys
  the environments of six damped Ly{\ensuremath{\alpha}} systems at z
  {\ensuremath{\approx}} 3},''
  \href{http://dx.doi.org/10.1093/mnras/stz1501}{{\em MNRAS} (Jun, 2019) 1435},
  \href{http://arxiv.org/abs/1904.07254}{{\ttfamily arXiv:1904.07254
  [astro-ph.GA]}}.

\bibitem{2012JCAP...11..059F}
A.~{Font-Ribera}, J.~{Miralda-Escud{\'e}}, E.~{Arnau}, B.~{Carithers}, K.-G.
  {Lee}, P.~{Noterdaeme}, I.~{P{\^a}ris}, P.~{Petitjean}, J.~{Rich},
  E.~{Rollinde}, N.~P. {Ross}, D.~P. {Schneider}, M.~{White}, and D.~G. {York},
  ``{The large-scale cross-correlation of Damped Lyman alpha systems with the
  Lyman alpha forest: first measurements from BOSS},''
  \href{http://dx.doi.org/10.1088/1475-7516/2012/11/059}{{\em JCAP} {\bfseries
  2012} no.~11, (Nov, 2012) 059},
  \href{http://arxiv.org/abs/1209.4596}{{\ttfamily arXiv:1209.4596
  [astro-ph.CO]}}.

\bibitem{1974ApJ...193..327P}
M.~{Peimbert} and S.~{Torres-Peimbert}, ``{Chemical composition of H II regions
  in the Large Magellanic Cloud and its cosmological implications.},''
  \href{http://dx.doi.org/10.1086/153166}{{\em Apj} {\bfseries 193} (Oct.,
  1974) 327--333}.

\bibitem{Li:2019zvm}
P.~Li, F.~Lelli, S.~McGaugh, M.~S. Pawlowski, M.~A. Zwaan, and J.~Schombert,
  ``{The halo mass function of late-type galaxies from HI kinematics},'' {\em
  Astrophys. J.} {\bfseries 886} no.~1, (2019) L11,
\href{http://arxiv.org/abs/1911.00517}{{\ttfamily arXiv:1911.00517
  [astro-ph.GA]}}.

\bibitem{Beltran:2005gr}
M.~Beltran, J.~Garcia-Bellido, J.~Lesgourgues, and M.~Viel, ``{Squeezing the
  window on isocurvature modes with the lyman-alpha forest},''
  \href{http://dx.doi.org/10.1103/PhysRevD.72.103515}{{\em Phys. Rev.}
  {\bfseries D72} (2005) 103515},
\href{http://arxiv.org/abs/astro-ph/0509209}{{\ttfamily arXiv:astro-ph/0509209
  [astro-ph]}}.

\bibitem{Sykes:2019sop}
C.~Sykes, M.~Fumagalli, R.~Cooke, and T.~Theuns, ``{Determining the primordial
  helium abundance and UV background using fluorescent emission in star-free
  dark matter haloes},''
\href{http://arxiv.org/abs/1912.06163}{{\ttfamily arXiv:1912.06163
  [astro-ph.CO]}}.

\end{thebibliography}\endgroup
\end{document}